\documentclass[letterpaper]{article}

\usepackage{prepr}

\title{Towards Integrating Emerging AI Applications in SE Education}
\author{
\PREPauthor{Michael Vierhauser}{University of Innsbruck, Department of Computer Science}{Michael.Vierhauser@uibk.ac.at}
\PREPauthor{Iris Groher}{Johannes Kepler University Linz, Institute of Business Informatics -- Software Engineering}{iris.groher@jku.at}
\PREPauthor{Tobias Antensteiner}{University of Innsbruck, Department of Computer Science}{tobias.antensteiner@uibk.ac.at}
\PREPauthor{Clemens Sauerwein}{University of Innsbruck, Department of Computer Science}{clemens.sauerwein@uibk.ac}
\vspace{0.6cm}
}
\setvenue{the \vspace{3pt}\textbf{\\36th Conference on Software Engineering Education and Training (CSEE\&T)}}
\setvenueone{the \textbf{36th Conference on Software Engineering Education\\ and Training (CSEE\&T)}}
\setyear{2024}
\setdoi{10.1109/CSEET62301.2024.10663045}

\begin{document}

\maketitle

 \begin{abstract}
Artificial Intelligence (AI) approaches have been incorporated into modern learning environments and software engineering (SE) courses and curricula for several years. However, with the significant rise in popularity of large language models (LLMs) in general, and \textit{OpenAI}'s LLM-powered chatbot ChatGPT in particular in the last year, educators are faced with rapidly changing classroom environments and disrupted teaching principles.
Examples range from programming assignment solutions that are fully generated via ChatGPT, to various forms of cheating during exams. 
However, despite these negative aspects and emerging challenges, AI tools in general, and LLM applications in particular, can also provide significant opportunities in a wide variety of SE courses, supporting both students and educators in meaningful ways.
In this early research paper, we present preliminary results of a systematic analysis of current trends in the area of AI, and how they can be integrated into university-level SE curricula, guidelines, and approaches to support both instructors and learners.
We collected both teaching and research papers and analyzed their potential usage in SE education, using the \textit{ACM} Computer Science Curriculum Guidelines CS2023.
As an initial outcome, we discuss a series of opportunities for AI applications and further research areas.
\end{abstract}

\section{Introduction}
\label{sec:intro}


Artificial Intelligence (AI) in education, particularly at the university level, has been an active area for several years now, ranging from intelligent tutoring systems to learning analytics support~\cite{salas2022artificial,le2013review}. 
However, the recent rise in popularity of large language models (LLMs) and chatbots has already disrupted teaching principles in many programming and software engineering (SE) courses. Solutions to programming exercises are automatically generated by students using, for example, ChatGPT and pose further challenges for educators when conducting online exams~\cite{rahman2023chatgpt}. 
Despite the challenges, AI tools in general, and LLM applications in particular, can also be meaningfully employed in a wide variety of SE courses, commonly part of computer science curricula, supporting learners in tasks such as test case generation, architecture pattern creation, or bug detection~\cite{schafer2023empirical,Xia2023}. 
Other application scenarios include providing individual feedback to students~\cite{Balse2023}, chatbots to increase learning experience~\cite{Benner2024}, or using AI-based tools as programming assistants~\cite{Arghavan2023}.
However, so far, only a fraction of these aspects and novel application areas are covered in SE education and curricula. Moreover, not only learners can benefit from these new opportunities, but also teachers and educators may use them, for example, for generating assignment exercises, or as part of automated grading and assessment systems~\cite{gonzalez2021artificial}. 
In either application scenario, however, the unreflected and careless use of these new tools can also impact the learning experience and outcomes in negative ways, posing new challenges to both students and educators~\cite{bergstrom2024challenges}. To exacerbate the situation, existing didactic concepts in the context of SE education hardly take into account or take advantage of these new tools. Also, universities often lack guidelines for the ethical and responsible usage of AI tools in education. Apart from that, students need to be prepared for the ethical implications and societal impacts of AI technologies.

In this paper, we present a preliminary analysis of current trends in the area of AI and how they can potentially be integrated into SE curricula.
As part of this, we specifically investigate how they can be leveraged in SE education for both students and educators and how didactic concepts can be adapted to appropriately deal with these new applications.
For this purpose, we (1) collected current research on this topic from the past years, (2) performed further snowballing, and (3) grouped the extracted approaches and application areas with regard to their potential usage in SE education, using the \textit{ACM} Computer Science Curriculum Guidelines version "Gamma"\footnote{Please visit \href{https://csed.acm.org/cs2023-gamma}{csed.acm.org/cs2023-gamma} for more information.}, currently under review.

The main goal of our ongoing work is to gain a better understanding of how the current state of research in AI applications is reflected in SE education, identifying  emerging challenges and opportunities for SE courses.
As a result, in this short paper, we present an initial overview of techniques and exemplary use cases, alongside a roadmap for further research on and applications of AI tools in SE education. 

\section{Study Design \& Related Work}
\label{sec:background}

In this section, we first provide a brief motivation for the need to systematically introduce AI in SE education, and further discuss existing related work in this area.

\subsection{Identifying AI Trends in Research and Education}

To gain better insights into current trends -- particularly within the past year -- we started a structured literature analysis for both research and educational venues in software engineering.
We specifically chose to include both research and education venues, hoping this will provide a more comprehensive understanding of current research trends and the extent to which they have been integrated into SE education.
Furthermore,  we aim to document challenges educators need to be aware of and, as a result, potentially adapt didactic concepts and/or teaching materials. In this short paper, we report on our work in progress and the first results of this analysis.
For this purpose, we discuss initial findings (cf.~\citesec{trends}) and potential applications and opportunities (cf.~\citesec{challops}). 

To enable a more structured analysis, we further aim to categorize our findings alongside the different SE activities and areas. For SE, numerous taxonomies and classification schemes have been proposed, most notably the Software Engineering Body of Knowledge (SEBoK)~\cite{swebok}, with its 18 knowledge areas.
As our primary focus, however, is on SE education, we have chosen to group our findings according to the ACM Computer Science Curriculum Guidelines version "Gamma" (CS2023), comprising 17 different knowledge areas, which also incorporate -- to a very large extent -- the areas proposed in the SEBoK.
The ACM Computer Science Curriculum Guidelines are designed to assist educators in the area of computer science, describing the general structure, as well as listing important topics to cover, core and elective CS courses~\cite{raj2022toward}.
Most notably, the most recent version, CS2023, currently discussed by the ACM task force, will take into account competency-based education, incorporating a competency model, comprising required skills, and knowledge to be acquired by students. Introducing Learning Analytics and competency-based education in classes typically requires a significant effort and is also something we see huge potential for leveraging AI and alleviating the burden of, for example, manually assessing if all teaching materials correspond to specified competencies (cf.~\citesec{challops}).

While CS2023 addresses a broad and comprehensive education, covering 17 different knowledge areas, ranging from computer science foundations, to AI, and parallel computing, we focus on two main areas: Software Development Fundamentals (\texttt{SDF}) that cover basic programming education, and second, Software Engineering (\texttt{SE}). A brief overview of the sub-categories for these two areas is  shown in~\citetable{results}.
The goal of our work is to identify existing work and approaches for each sub-category of the two areas and collect different means of how AI can be incorporated, for example in teaching aspects of software architecture, such as design patterns, or architectural styles. 
For this purpose, as a starting point, to gain an initial understanding of current research and trends, we collected an initial set of research papers from 2023, of two prime software engineering conferences: the International Conf. of Software Engineering
, and the International Conf. on the Foundations of Software Engineering
, as well as four educational conferences/tracks: Software Engineering Education and Training Track@ICSE, the International Conf. on Software Engineering Education and Training
, the Hawaii International Conf. on System Sciences, 
 and the Conf. on Innovation and Technology in Computer Science Education.

The goal was to identify trends and hot topics and uncover areas that might be covered by research but not yet as part of the CS and SE education. In total, we started with a set of 51 publications (33 research-related and 18 education-related), which was extended to 71 after performing an extra round of snowballing.
We focused on publications that presented AI applications or educational concepts in the area of SE and programming, but excluded any kind of study, analysis, or review that did not present a specific approach. After carefully reading the titles and abstracts, we excluded 17 papers that were deemed out of scope (e.g. tertiary studies), with a final set of 54 papers remaining.

After selecting our set of papers, we divided them randomly, with four researchers evaluating the papers independently. For the research papers, we focused on the purpose (e.g., test case generation), the technology used, and the specific application area. (Please note, that for the purpose of this paper, and for the sake of brevity, we mainly focus on the intent, and the relation to the CS2023 categories, but future work will broaden the scope of this analysis). For education papers, we also focused on the CS2023 categories, and tried to relate them to the research areas -- and ultimately identify gaps and areas that are not yet considered in education.
Based on this analysis, we then selected the respective ACM curriculum categories (i.e., Knowledge Units) where the proposed approach, tool, or research could be relevant or applied. After this step, we consolidated all assessments, discussed and resolved conflicts until mutual agreement was achieved, and tried to identify clusters of tools and approaches, as well as gaps.
The final goal of the study is to derive a set of concrete actionable guidelines for different courses part of the CS curriculum, with a set of tools and approaches that could be used, and aspects that need to be considered. In the following sections, we discuss the initial assignment of approaches to the CS2023 Knowledge Units and potential application scenarios.

\subsection{Related Work}
Several systematic reviews have investigated and classified aspects of AI applications in education.
Bittencourt~\etal[bittencourt2023positive] conducted an analysis exploring the intersection of positive psychology and AI in education (P-AIED), identifying P-AIED as a new global movement focusing on positive emotion and engagement related to AI-supported teaching and learning.
Kuhail~\etal[kuhail2023interacting] and Okonkwo and Ade-Ibijola~\cite{okonkwo2021} performed studies related to the application of chatbots in education. Kuhail~\etal analyzed work on educational chatbots and found that these are primarily used in computer science, language learning, general education, and to a lesser extent in fields like engineering and mathematics. Most chatbots were web-based and more than half functioned as teaching agents, while over a third acted as peer agents. Okonkwo and Ade-Ibijola analyze 53 primary studies and identify an increasing trend in the application of chatbots in education. 

Zhai~\etal[zhai2021review] analyzed 100 papers focusing on the application of AI in the education sector from 2010 to 2020. They classified primary studies into a development layer (classification, matching, recommendation, and deep learning), an application layer (feedback, reasoning, and adaptive learning), and an integration layer (affection computing, role-playing, immersive learning, and gamification).
Chen~\etal[chen2020] performed a study investigating the impact of AI on education, focusing on its application and effects in administrative, instructional, and learning contexts. 
Similarly, Zawacki-Richter~\etal[zawacki2019systematic] conducted a systematic review of research on the application of AI in higher education. The majority of research comes from the fields of Computer Science and STEM with AI being mainly used in areas of student profiling and predictive analytics, as well as intelligent tutoring systems.

Several studies have focused on AI in computer science education.
Denny~\etal[denny2023computing] explore the challenges and opportunities presented by recent advances in AI, particularly code generation models, and their potential impact on computing education. 
Kalles~\cite{Kalles2016} reports on the experience of using AI systems as the basis of educating IT students. The application areas were mainly decision tree lifecycle management and board game learning mechanisms.
Daun and Brings~\cite{p037} discuss the role of generative AI technologies, such as ChatGPT, in SE education. They discuss potential risks associated with students' use of generative AI but also identify several opportunities that these technologies can offer in enhancing educational practices such as individualization of education and personalized feedback.

Current work in this regard primarily addresses only parts of SE education, for example, programming. With our work, we aim to cover a broader spectrum of SE education, leveraging the  ACM Computer Science Curriculum Guidelines.

\section{Current AI Trends \& Applications}
\label{sec:trends}

\begin{table*}[]
    \footnotesize
    \centering
    \vspace{0.3cm}
    \renewcommand{\arraystretch}{1.15}

     \addtolength{\tabcolsep}{-2.0pt}
     \caption{ACM CS2023 categories of the two selected knowledge areas \texttt{SDF} and \texttt{SE}~(cf.~\citesec{background}) with examples from our literature search and coverage in research/education (\CIRCLE\,=\,Covered in several publications; \RIGHTcircle\,=\,Partially Covered; \Circle\,=\,Not/Sparesly Covered), and the number of papers (\#) with research (R) and education (E) focus.}
    \label{tab:results}
   \begin{tabular}{L{0.9cm}L{4.2cm} L{10.385cm}R{1.1cm}p{0.6cm}}
     \toprule
{\bf ID}&{\bf Category}&{\bf Description/Example} & {\bf \# (R/E)}  
 & {\bf Co.}\\ \midrule

             \multicolumn{5}{c}{\emph{Software Development Fundamentals (\texttt{SDF})}}\\\midrule

\texttt{SDF\,1} & Fundamental Programming Concepts and Practices &
\multirow{3}{\linewidth}{Basic programming concepts, in conjunction with algprithms and data structures, can be supported by a variety of emerging approaches. 
Particularly, LLMs can be employed for many different tasks and support roles such as the generation of programming assignments~\cite{p036}, serving as chatbots and interactive tutors~\cite{p037}.}
& 16 (2/14) & \CIRCLE \\
\texttt{SDF\,2} & Fundamental Data Structures & & 13 (0/13) &\CIRCLE \\
\texttt{SDF\,3} & Algorithms & & 11 (0/11) & \CIRCLE\\
\texttt{SDF\,4} & Software Development Practices & 
AI approaches can guide and improve development practices, e.g., by generating or analyzing commit messages~\cite{p011} or explaining code snippets~\cite{p039}.
& 8 (5/3) & \RIGHTcircle\\

                \midrule
                 \multicolumn{5}{c}{\emph{Software Engineering (\texttt{SE})}}\\\midrule

\texttt{SE\,1} & Teamwork & AI approaches supporting collaborative work, e.g., for generating commit messages~\cite{p010}.  &1 (1/0) &\Circle \\
\texttt{SE\,2} & Tools and Environments & AI-supported SE tools, e.g., chatbots supporting novice developers~\cite{p045}.& 9 (6/3)& \RIGHTcircle\\
\texttt{SE\,3} & Product Requirements & AI-supported question answering tools help analyze and understand NL requirements~\cite{ezzini2023ai}. & 1 (1/0)& \Circle\\
\texttt{SE\,4} & Software Design & \multirow{2}{\linewidth}{AI techniques, including knowledge extraction for APIs~\cite{p020}, code generation~\cite{p023}, and advances code search accuracy~\cite{p024}, and systems like AlphaCode~\cite{p053}.}
& 2 (2/0)& \Circle\\
\texttt{SE\,5} & Software Construction & &17 (11/6) & \CIRCLE\\
\texttt{SE\,6} & Software Verification and Validation & Novel tools leveraging AI approaches to detect bugs, e.g., using CodeBERT~\cite{p002}.& 24 (22/2) & \CIRCLE\\
\texttt{SE\,7} & Refactoring and Code Evolution &  
AI techniques for code evolution, e.g., for multi-language applications/systems~\cite{p003}. & 9 (9/0)
& \RIGHTcircle  \\
\texttt{SE\,8} & Software Reliability &  ML \& NLP methods for code translation, inconsistency detection, and remediation. & 20 (19/1) & \CIRCLE\\
\texttt{SE\,9} & Formal Methods & While only sparsely covered so far, LLMs can also be used for, e.g. proof-generation~\cite{p001}. &1 (1/0) & \Circle\\

\bottomrule
    \end{tabular}
\vspace{-2pt}
\end{table*}

The classification of the 54 recent AI papers provides several insights into potential areas where AI topics can be incorporated into computer science and SE courses.
\citetable{results} provides an overview of the curriculum categories and number of papers relevant to this category, alongside some examples.
In the following, we provide a brief summary\footnote{Please refer to \href{https://doi.org/10.5281/zenodo.11316028}{doi.org/10.5281/zenodo.11316028} for the full classification.} of potential applications in four distinct categories, which we identified by grouping the categories with the highest coverage.

\paragraph{Introductory Programming Education (cf. \texttt{SDF\,1} to \texttt{SDF\,4})} The integration of AI into courses for novice programmers offers novel methods for enhancing the learning experience and addressing educational challenges. Recent works have studied the application of LLMs to block-based languages~\cite{Griebl2023} which are often used in introductory programming courses. Incorporating AI-based code generation tools~\cite{Reeves2023} can offer real-time coding assistance, thereby facilitating a deeper and more rapid comprehension of programming fundamentals. Furthermore, pair programming assistants~\cite{Arghavan2023} can provide on-demand, context-aware coding suggestions.

\paragraph{Design and Construction  (cf. \texttt{SDF\,4},  \texttt{SE\,1}, \texttt{SE\,3} to \texttt{SE\,5})} Given that software design focuses on the engineering-oriented design of internal software components, "machine-learning-based knowledge extraction techniques"~\cite{p020} can be employed not only to discover, but also to better understand application programming interfaces (APIs).
In the realm of software construction, recent studies have shed light on the robustness of code generation techniques, such as \textit{GitHub Copilot}, which automatically generates code from natural language descriptions~\cite{p023}. In addition, advances in so-called contrastive learning, exemplified by \textit{CoCoSoDa}, have significantly enhanced the accuracy and performance of code search~\cite{p024}. Moreover, the emergence of systems such as \textit{AlphaCode} represent a breakthrough as they demonstrate competitive-level code generation capabilities, impacting problem-solving in programming competitions~\cite{p053}.

\paragraph{Maintenance and Testing (cf. \texttt{SE\,2}, \texttt{SE\,7})}
The integration of AI into more advanced SE courses is currently only sparsely covered in education research, despite the existence of several research advances in this area. This includes, for example, automated test case generation or test augmentation support~\cite{nie2023learning}. LLMs have also been employed in this context ~\cite{kang2023large} fed with bug reports and tasked with generating a series of test candidates. Incorporating these tools in advanced software engineering, and software testing courses can foster awareness of these new tools and technologies. It is, however, also important to note, and convey the message, that these can only serve as an augmentation to existing, traditional, testing methods and are not meant to replace them. 

\paragraph{V\&V, Reliability, and Security (cf. \texttt{SE\,6}, \texttt{SE\,8}, \texttt{SE\,9})}
The integration of new tools using AI approaches for software verification and validation to recognize bugs or identify vulnerabilities has become an established research area. For example, models such as \textit{CodeBERT} are used to recognize errors in source code~\cite{p002}. In addition, the combination of ML and NLP has led to the development of novel AI approaches enabling code translation~\cite{p003}, code repair, and inconsistency checking~\cite{p011} to increase reliability and security of code.


\section{Challenges \& Opportunities in SE Education}
\label{sec:challops}

Based on the analysis of trending topics and educational use, we have identified several areas of opportunities where AI could support both educators and students in various tasks. 
In the following section, we present two directions currently not -- or only partially -- covered in SE education research, where (1) AI tools should be considered, how AI tools and methods could be applied to (2) alleviate exercise creation as well as grading tasks, and (3) provide tailored support for students.
Our literature search has shown that current work related to AI in SE education has a strong focus on exercise generation~\cite{speth2023investigating}, assessment~\cite{gonzalez2021artificial}, and grading~\cite{Balse2023} using LLMs. However, the quality of the results varies and often does not yield satisfactory programming tasks or grading results (cf. \hyperref[para:direction_1]{Direction~I}). Furthermore, going beyond introduction to programming courses, i.e., CS1, as well as basic data structures courses, i.e., CS2, we observed, that while being a very active research topic, only a few papers address more advanced areas in SE education (cf. \hyperref[para:direction_1]{Direction~II}). 


\paragraph{Direction I -- Competency-focused Exercise Generation (Educators)}\label{para:direction_1} Following the trend of employing generative AI for a slew of different tasks, one potential application area could be the creation of more diverse and comprehensive exercises, e.g., for programming classes. While similar works have been proposed recently~\cite{speth2023investigating,freitas2023nlp}, we think that this can go far beyond "simply" creating exercise sheets or programming tasks.
Particularly taking into account competency-based education~\cite{pluff2022competency}, which is also reflected in the most recent 2023 ACM curriculum guidelines, with a dedicated competency model, LLMs could be used in this context. First, to create a set of tasks that covers certain competencies, e.g., specified in a competency model, but also to generate specific test cases that cover/check the achievement of certain competencies, and even provide support for assigning assignments and identifying gaps in missing competences or tasks that exceed what students should be able to accomplish as part of an assignment.

\emph{The challenge} hereby resides in providing the LLM with the necessary information -- e.g., a competency model -- and the context and scope of the assignment and tasks.

\emph{The benefits} are on the one hand a significant reduction in the manual tedious effort of creating variations of exercises (e.g., programming tasks) and on the other hand a move towards competency-based assessment and Learning Analytics support, without the need to manually incorporate these aspects in exercises and assignments.

\paragraph{Direction II -- AI in SE (Educators and Students)}
Various authors have stressed the application of AI tools in Software Development Fundamentals (\texttt{SDF\,1}, \texttt{SDF\,2}, and \texttt{SDF\,3}), both as an opportunity and a challenge (e.g., students generating assignments without critical reflection). However, the deeper application of AI tools in advanced SE classes does not seem to be widely considered~\cite{johnson2024}.
In contrast, we discovered a wide variety of AI applications in research that intersect with basic and advanced software engineering education. Particularly software quality assurance, maintenance, and validation (\texttt{SE\,6} and \texttt{SE\,8}) are covered by several research papers, but are hardly present in the educational context.

\emph{The challenge} from our perspective is striking a balance between those new technologies and tools that alleviate certain tasks and still teaching basic and advanced concepts and theories to students. For example, if code generation tools take over tasks, such as generating architectural patterns or unit tests, it is still crucial to convey knowledge about these foundational aspects so that students know how to apply them properly and understand the concepts behind them. 

\emph{The benefits} are that students are introduced to new tools and paradigms early on in their education that they are very likely to encounter later on during their professional careers. Moreover, it fosters understanding of AI capabilities and limitations, preparing students to understand ethical implications and societal impacts with a well-informed, critical perspective.

\section{Outlook and Conclusion}
\label{sec:conclusion}

With the release of ChatGPT, AI tools, and particularly LLMs have significantly impacted both educators and students, often being perceived as a boon and a bane at the same time.
While providing a vast amount of opportunities for new and improved tools for generating code, analyzing bugs, or improving code quality, these aspects are only sparsely converted in current software engineering courses.
In this early research paper, we present a preliminary analysis of current trends in the area of AI, and how they can be integrated into SE curricula. We further discuss several challenges and opportunities where AI-powered tools and techniques can support teachers and learners, but also need to be taken into account when designing exercises or course assignments.

We will further explore the development and evaluation of didactic methods that integrate AI tools into university-level courses. Additionally, there is a need to investigate the long-term impacts of AI integration on learning outcomes and employability of students.

\balance
\bibliographystyle{abbrv}
\bibliography{refs}

\end{document}